\documentclass[11pt]{article}
\usepackage{amsmath}
\usepackage{amsfonts}
\usepackage{amssymb}
\usepackage{graphicx}

\def\be\begin{equation}
 \def\ee{\end{equation}}
\def\bea{\begin{eqnarray}}
\def\eea{\end{eqnarray}}

\begin{document}
\begin{center}
\LARGE {\bf Bouncing cosmology without anisotropy  }
\end{center}
\begin{center}
{\bf  Kh. Saaidi}\footnote{ksaaidi@uok.ac.ir},
{\bf  S. Ziaee}\footnote{S.ziaee@uok.ac.ir}\\

{\it Department of Physics, Faculty of Science, University of
Kurdistan,  Sanandaj, Iran}

\end{center}
 \vskip 1cm
\begin{center}
{\bf{Abstract}}
\end{center}
Using non-linear equation of state for pressure and density
energy, we show that the universe is began with a smooth and
isotropic bounce. We use a non-linear equation of state which is
a binary mixture of perfect fluid and dark energy. We show that in
order to preserve a smooth and isotropic bounce, the source for
contraction before the bounce, must have an equation of state
with $\omega>1$ (Ekpyrotic matter) and a dark energy with
positive pressure.

\newpage

\section{Introductions}

The standard cosmological model furnishes an accurate description
of the evolution of the universe, since 14 billion years ago
approximately {\cite{1}}. Regardless of its success, the standard
cosmological model suffers from a few problems such as the initial
singularity, the cosmological horizon, the flatness problem, the
baryon asymmetry, the homogeneity problem, the large scale
structure problem, and the nature of dark matter and dark energy
{\cite{2}}. Although inflation partially answer some problem, it
does not solve the essential problem of the initial singularity
{\cite{3}}. The existence of an initial singularity is perturbing.
Singularity can be naturally considered as a source of
lawlessness {\cite{4}}, because the spacetime explanation breaks
down there, and physical laws presuppose spacetime.
\par In the begining of the 80's, it was clear that the standard cosmological
model was in crisis. The existence of the inflationary theory gave
an answer to some of these problems and opened the window for a
description of the origin of the spectrum of primeval
fluctuations. In fact, inflation predicts the appearance of
quantum fluctuations in the initial vacuums state, arrives to
primeval perturbations seeding the observed cosmic large-scale
structures {\cite{5}}. These primeval fluctuation are endowed with
a nearly scale-invariant spectrum, in agreement with observation
of the cosmic microwave background {\cite{6}}. It is well know
that inflation produce an explanation for the homogeneity,
flatness and horizon problem of the standard hot Big Bang
cosmology. However, in spite of its successes, the theory of
inflation does not solve the problems of the initial singularity
and can not embedding inflation within a quantum gravity.\\
Mainly inspired by the string motivated pre-Big Bang scenarios
{\cite{7,8}}, bouncing models {\cite{9,10,11,12,13}}, i.e. models
in which the universe undergoes a phase of contraction followed by
expansion, have been proposed as alternatives to the inflationary
paradigm {\cite{14}}.
\par The difficulties of embedding inflation within a quantum gravity
theory and the persistence of the initial singularity in the
inflationary scenario have motivated several proposals of
alternative cosmologies. There is a general consensus on the
existence of a high energy
 cut-off at the order of the Planck scale, at which classical general relativity should be replaced
by a quantum gravity theory. From this point of view, the Big
Bang singularity just represents the outcome of the extrapolation
of general relativity beyond its domain of applicability, whereas
the quantum gravity theory should regularize this singularity,
replacing it by a maximum in the curvature and energy density of
the universe. The existence of a contraction phase before the Big
Bang has been argued in several frameworks. Following this
hypothesis, the universe should contract from initial conditions
in a low energy regime, evolving into a phase of higher and
higher curvature, until the high energy cut-off of the true
quantum gravity theory comes into play. This reverses the
contraction into a standard decelerated expansion, thus avoiding
the general relativistic singularity and replacing it by a cosmic
bounce {\cite{24}}.
\par Bouncing cosmology model can simply solve the problems of
flatness and horizon from standard cosmology model, but
anisotropy in contraction phase is troublesome when the
contribution phase is immaterial, anisotropy become quickly
dominant and leads to a $velocity$ $dominance$ $singularity$
{\cite{15,16}}. This typical result of general relativity can
only be avoided if energy density of  matter source growth more
quickly than anisotropy. On the other hand, all sources with
$\omega$ smaller than one growth very slowly and at the end they
become anisotrop. When this occur, a mixmaster theory takes place
with the development of chaotic $Belinskii$, $Khalatnikov$ and
$Lifshitz$ ($BKL$) oscillations in the scale factors
{\cite{15,16,17,18}}. Therefore, there is a probability that
cross universe after bounce is unreliable anisotrop. According to
this point that when strong curvature effect is dominant,
mixmaster treatment takes place in high energy phase, so they rise
this probability to solve anisotropy in contraction phase problem
by adding a
non-linear term to equation of state (EoS) [24].\\
In this paper we investigated the effects of a general non-linear
term of EoS in the bouncing cosmological model. In order to
understand this context the non-linear EoS can isotopize the
universe at early times and at high energy regime, when the
bounce is approached. We wanted to study on the possible use of a
non-linear EoS as an effective way of representing a dark energy,
to solve the anisotropy problem in contraction phase. Also, here
we obtain the density energy, $\rho$, and anisotropic factor,
$\sigma^2$, for case that the EoS is a binary mixture of perfect
fluid and dark energy.

\section{Non-linear EoS}

We assume that gravity in the contraction regime is determinate
by Einstein equation. And also we suppose that the contraction
regime is dominated by a binary mixture of a perfect fluid and
dark energy with energy density $\rho$ and pressure $P$. By
$P_m=\omega\rho$ and a dark energy component with non-linear form
of $\rho$. We consider general form of a non-linear EoS as follows
\begin{equation}
P=P_m+P_d=\omega\rho+\epsilon\frac{\rho^\alpha}{\rho_c^{(\alpha-1)}},
\end{equation}
where $\omega$ is a constant value (is a pure number), that
indicates the low energy EoS of the fluid, $\rho_c>0$ is the
transition scale and $\epsilon$ is the sign of the non-linear
term. Here we will alone focus on the case $\epsilon>0$, where it
means that the pressure of dark energy is positive.
\par The easiest way to investigate the behavior of anisotropy in the
contraction phase, is using from Bianchi type I models. The
Bianchi type I models are a subclass of the Bianchi class A
models. These models are homogeneous and anisotropic cosmological
models including the flat Fridmann model {\cite{19}}. The Bianchi
type I cosmology can be defined by the Hubble expansion scaler
and the tracefree shear tensor $\sigma_{\mu\nu}$, in which
$\mu,\nu=1,...,3$ and $\sigma^2= \frac{1}{2} \sigma_{\mu\nu}
\sigma^{\mu \nu }$. The energy-momentum tensor is give by
\begin{equation}
T_{\mu\nu}=(\rho+P)u_{\mu}u_{\nu}-Pg_{\mu\nu},
\end{equation}
where $\rho$ is the energy density, $P$ the pressure and $u$ is
the 4-vector velocity. \\
The energy conservation equation for a
cosmological model including a perfect fluid is
\begin{equation}
\dot{\rho}=-3H(\rho+P),
\end{equation}
where $H$ is the Hubble parameter, i.e. $H=\frac{\dot{a}}{a}$.
Using the Einstein equations and the Bianchi type I models with
assume $\frac{8{\pi}G}{c^4}=1$, we can obtain

\begin{equation}
H^2=\frac{1}{3}(\rho+\sigma^2),
\end{equation}

\begin{equation}
\dot{H}=-H^2-\frac{1}{6}\left((\rho+3P)+4\sigma^2\right),
\end{equation}

\begin{equation}
\dot{\sigma}=-3H\sigma.\\
\end{equation}
The energy conservation (3) and non-linear EoS (1) can determine
the energy density as a function of scale factor.
\begin{equation}
\rho=\rho_c\left[
\frac{(1+\omega)A^{(\alpha-1)}}{a^{3(1+\omega)(\alpha-1)}-\epsilon
A^{(\alpha-1)}} \right]^{\frac{1}{\alpha - 1}},
\end{equation}
\begin{equation}
A=\frac{{\rho_0}a_0^{3(1+\omega)}}{[(1+\omega)\rho_c^{(\alpha-1)}+\epsilon{\rho_0^{(\alpha-1)}}]^\frac{1}{\alpha-1}},\\
\end{equation}
where $\rho_0$  and  $a_0$ express the energy density and scale
factor at an arbitrary time $t_0$. This is acceptable for all
values of $\epsilon$, $\rho_0$ and $\omega$ except for
$\omega=-1$ {\cite{19}}. Defining $a_*=|A|^\frac{1}{3(1+\omega)}$
and supposing $\rho>0$, we arrive at
\begin{equation}
\rho=\rho_c\left[\frac{(1+\omega)}{(\frac{a}{a_*})^{3(1+\omega)(\alpha-1)}-\epsilon}\right]^{\frac{1}{\alpha-1}}.\\
\end{equation}
Using Eq.(6), we can obtain shear scales as a function of the
scale factor as
\begin{equation}
\sigma^2=\sigma_i^{2} \left( \frac{a}{a_i} \right)^{-6}.\\
\end{equation}
According to above equation, the shear growing larger for smaller
$a$ in the past and quickly decays at late times. \\Note that in a
contraction phase driven by a source fulfilling the energy
conditions, hence we limited our work to the state $(1+\omega)>0$.
\\The quantity of $a_*$ can be obtained from the initial
conditions of the universe, so if the universe begins with a scale
factor $a_i$ we arrive at
\begin{equation}
a_*=a_i \left [ (1+\omega) \left(\frac{\rho_c}{\rho_i}
\right)^{(\alpha-1)}+\epsilon
\right]^{-\frac{1}{3(1+\omega)(\alpha-1)}}.\\
\end{equation}
\par As a result of the existence of a contraction phase before the
Big Bang that has been discussed in several articles
{\cite{20,21,22,23}}. Our universe contract from initial
conditions in a low energy phase, then larger and larger energy
scales until it get to planck scale. In this case, the existence
of a high energy cut-off at the order of the Planck scale and
classical general relativity replaced by a quantum gravity
theory. Therefore, when $\rho$ reaches $\rho_M$ quantum gravity
theory comes into play and in this stage the contraction is
stopped and the universe goes towards a standard decelerated
expansion phase from a bounce. So that we defined the scale
factor of the universe in the beginning of the bounce as
\begin{equation}
a_M=a_* \left [ (1+\omega) \left(\frac{\rho_c}{\rho_M}
\right)^{(\alpha-1)}+\epsilon
\right]^\frac{1}{3(1+\omega)(\alpha-1)}.\\
\end{equation}
\par One can estimate the anisotropic behavior of the universe
from comparing the contribution of shear term with respect to
matter term in Eq.(4), where $\sigma$ term caused by space
anisotropy. Then to avoiding an anisotropy approach to the bounce
this term should be very smaller then the matter term at the
onset the bounce. Therefore, using Eqs.(9-12), we can arrive at
\begin{equation}
\frac{\sigma_M^2}{\rho_M}=\frac{\sigma_i^2}{\rho_M}\left[
\frac{\left(\frac{\rho_c}{\rho_i}
\right)^{(\alpha-1)}+\frac{\epsilon}{1+\omega}}{\left(\frac{\rho_c}{\rho_M}
\right)^{(\alpha-1)}+\frac{\epsilon}{1+\omega}}
\right]^{\frac{2}{(1+\omega)(\alpha-1)}}.\\
\end{equation}

\section{Typical example}

\par In this section we want study some example which EoS of them
are non-linear for considering the anisotropic behavior of the
universe.

\subsection{The EoS with a quadratic term}

For $\alpha=2$, Eq.(1) shows the EoS with a quadratic term as
\begin{equation}
P=\omega\rho+\epsilon{\frac{\rho^2}{\rho_c}}.
\end{equation}
this form of non-linear EoS is steadied in {\cite{24}}. By the
way, from Eqs.(9,11,12), we have
\begin{equation}
\rho={\frac{(1+\omega)\rho_c}{(\frac{a}{a_*})^{3(1+\omega)}-\epsilon}},
\end{equation}
\begin{equation}
a_*=a_i \left [ (1+\omega) \left(\frac{\rho_c}{\rho_i}
\right)+\epsilon \right]^{-\frac{1}{3(1+\omega)}},
\end{equation}
\begin{equation}
a_M=a_* \left [ (1+\omega) \left(\frac{\rho_c}{\rho_M}
\right)+\epsilon \right]^\frac{1}{3(1+\omega)}.\\
\end{equation}
In this case, according to Eq.(15) to satisfy the assumption
$\rho>0$, a should satisfy $a_*<a<\infty$. Using Eqs.(15-17) and
(13) with imposing the hierarchy $\rho_M\gg\rho_c\gg\rho_i$ for
anisotropy fraction, we can obtain
\begin{equation}
\frac{\sigma_M^2}{\rho_M}\simeq
\frac{\sigma_i^2}{\rho_i}\left(\frac{\rho_c}{\rho_i}\right)^{\frac{1-\omega}{1+\omega}}{\left(\frac{\rho_c}{\rho_M}\right)}.\\
\end{equation}
\par Eqs.(15-18) show that these result is in agreement
with the obtained results in {\cite{24}}.

\subsection{Modified Polytropic Like Gas}

For $\alpha=1+\frac{1}{n}$, Eq.(1) shows the EoS of modified
polytropic like gas, where $n$ index is a positive ($n>0$). So,
from Eq.(9) we have
\begin{equation}
\rho=\rho_c\left[\frac{(1+\omega)}{(\frac{a}{a_*})^{\frac{3}{n}(1+\omega)}-\epsilon}\right]^{n},
\end{equation}
also, from Eqs.(11,12) we have
\begin{equation}
a_*=a_i \left [ (1+\omega) \left(\frac{\rho_c}{\rho_i}
\right)^{(\frac{1}{n})}+\epsilon \right]^{-\frac{n}{3(1+\omega)}},
\end{equation}

\begin{equation}
a_M=a_* \left [ (1+\omega) \left(\frac{\rho_c}{\rho_M}
\right)^{(\frac{1}{n})}+\epsilon \right]^\frac{n}{3(1+\omega)}.\\
\end{equation}
In this case, according to Eq.(19) and to satisfy the assumption
$\rho>0$, a should satisfy $a_*<a<\infty$.
\par Now we like consider anisotropy behavior for the EoS of
modified polytropic like gas. Using Eqs.(19-21) and (13) with
imposing the hierarchy $\rho_M\gg\rho_c\gg\rho_i$, we can obtain
\begin{equation}
\frac{\sigma_M^2}{\rho_M}\simeq
\frac{\sigma_i^2}{\rho_i}\left(\frac{\rho_c}{\rho_i}\right)^{\frac{1-\omega}{1+\omega}}{\left(\frac{\rho_c}{\rho_M}\right)},
\end{equation}
where $\frac{\sigma_i^2}{\rho_i}$ is initial anisotropy,
$\left(\frac{\rho_c}{\rho_i}\right)^{\frac{1-\omega}{1+\omega}}$
is growth factor that is arising from low energy phase, and
${\frac{\rho_c}{\rho_M}}$ is a reducing factor of anisotropy that
is arising from high energy phase. According to Eq.(22) growth
factor of anisotropy depends on the $\omega$ index that is the
linear term coefficient in the EoS. When $\omega$ is more than
one $(\omega>1)$, the growth factor is transformed to an
additional reducing factor. As for Ekpyrotic matter (super-stiff
matter), is $\omega>1$.
\par By transition scale $\rho_c$ is determined the efficiency of
growth factor in the linear term and reducing factor in the
non-linear term of EoS. If $\rho$ value is very close to
$\rho_c$, then just the growth factor caused by the linear term
remains. While if $\rho_c$ is very close to $\rho_i$, then only
the reducing factor caused by the non-linear term remains.
\subsection{Modified Chaplygin Like Gas}

For $-1<\alpha<0$, Eq.(1) shows the EoS of modified chaplygin like
gas. So, from Eq.(9) we have
\begin{equation}
\rho=\rho_c\left[\frac{(\frac{a_*}{a})^{3(1+\omega)(1-\alpha)}-\epsilon}{(1+\omega)}\right]^{\frac{1}{1-\alpha}},
\end{equation}
Also from Eq.(11,12) we have
\begin{equation}
a_*=a_i \left [ (1+\omega) \left(\frac{\rho_i}{\rho_c}
\right)^{(1-\alpha)}+\epsilon
\right]^{\frac{1}{3(1+\omega)(1-\alpha)}},
\end{equation}
\begin{equation}
a_M=a_* \left [ (1+\omega) \left(\frac{\rho_M}{\rho_c}
\right)^{(1-\alpha)}+\epsilon
\right]^{-\frac{1}{3(1+\omega)(1-\alpha)}}.\\
\end{equation}
In this case, according to Eq.(23) to satisfy the assumption
$\rho>0$, a should satisfy $0<a<a_*$. Using Eqs.(23-25) and (13)
 with imposing the hierarchy $\rho_M\gg\rho_c\gg\rho_i$, we can
obtain
\begin{equation}
\frac{\sigma_M^2}{\rho_M}\simeq
\frac{\sigma_i^2}{\rho_i}\left(\frac{\rho_M}{\rho_c}\right)^{\frac{1-\omega}{1+\omega}}{\left(\frac{\rho_i}{\rho_c}\right)},
\end{equation}
where $\frac{\sigma_i^2}{\rho_i}$ is initial anisotropy,
$\left(\frac{\rho_M}{\rho_c}\right)^{\frac{1-\omega}{1+\omega}}$
is growth factor that is arising from high energy phase, and
$\frac{\rho_i}{\rho_c}$ is a reducing factor of anisotropy that
is arising from low energy phase. Similarly with the case before,
the growth factor of anisotropy depends on the $\omega$.
Therefore, for $\omega>1$ the growth factor is transformed to an
additional reducing factor (in additional to
$\frac{\rho_i}{\rho_c}$ term).
\par Note that in this case, if $\rho_c$ is very close to the
bounce scale $\rho_M$, then the growth factor shrinks to one and
alone reducing factor caused by the linear term remains. While if
transition scale $\rho_c$ is very close to $\rho_i$, then the
reducing factor disappears and alone the growth factor caused by
the non-linear term of EoS remains.
\par Considering the Eqs.(18,22,26) the behavior of anisotropy surely depends on the initial amount of anisotropy
$\frac{\sigma_i^2}{\rho_i}$ in the initial conditions. According
to Eq.(18,22) if the amount of $\frac{\sigma_i^2}{\rho_i}$ is
sufficiently low, we can have $\rho_c$ relatively close to
$\rho_M$ then the effect of non-linear term is reduced. On the
other hand, if the value of initial anisotropy is too high, then
$\rho_c$ should be very close to $\rho_i$. Therefore, the has an
important role in reducing the anisotropy in this case. According
to Eq.(26), if the universe is previously fairly isotropic, we can
have amount of $\rho_c$ relatively close to $\rho_i$. Instead, if
the universe begins in a very anisotropic case, therefore, for
reducing the anisotropy, we should be take $\rho_c$ very close to
$\rho_M$, i.e. the effect of non-linear term in the EoS decreases.
\par In general, according to resulted equations, to preserve a smooth and isotropic bounce, the source
of contraction should be a EoS with $\omega>1$. Considering that
linear EoS can not lonely solve the problem of anisotropic. Thus
by addition general non-linear term to EoS and resulting a similar
equations to (22) and (26) for $\omega>1$ we can solve anisotropy
problem in contraction phase. In fact, in the case of
$ekpyrotic/cyclic$ and $pre-Big$ $Bang$ models the initial
expansion is only isotropic if $\omega>1$ as in the case of
general relativity {\cite{25}}.

\section{conclusion}
In this work, we have studied the early time behavior of
anisotropy in contraction phase close to the bounce. Here we
introduce a general non-linear EoS and investigated the behavior
anisotropy of universe at early times and at high energy regime.
Specially we solved some typical example and we have shown that
to a smooth and isotropic bounce we must have a Ekpyrotic
$(\omega>1)$ matter with a dark energy component with positive
pressure at the onset of the bounce, which we called them
generalized chaplygin like gas and polytropic like gas.

\end{document}